\documentclass[twocolumn,showpacs,preprintnumbers,amsmath,amssymb]{revtex4}

\usepackage{graphicx}
\usepackage{dcolumn}
\usepackage{bm}

\begin{document}


\title{Topological Superconductivity, Topological Confinement \\ and the Vortex Quantum Hall Effect}

\author{M. Cristina Diamantini}
\email{cristina.diamantini@pg.infn.it}
\affiliation{%
INFN and Dipartimento di Fisica, University of Perugia, via A. Pascoli, I-06100 Perugia, Italy
}%


\author{Carlo A. Trugenberger}
\email{ca.trugenberger@InfoCodex.com}
\affiliation{%
SwissScientific, chemin Diodati 10, CH-1223 Cologny, Switzerland
}%


\date{\today}

\begin{abstract}
Topological matter is characterized by the presence of a topological BF term in its long-distance effective action. Topological defects due to the compactness of the U(1) gauge fields induce quantum phase transitions between topological insulators, topological superconductors and topological confinement. In conventional superconductivity, due to spontaneous symmetry breaking, the photon acquires a mass due to the Anderson-Higgs mechanism. In this paper we derive the corresponding effective actions for the electromagnetic field in topological superconductors and topological confinement phases. In topological superconductors magnetic flux is confined and the photon acquires a topological mass through the BF mechanism: no symmetry breaking is involved, the ground state has topological order and the transition is induced by quantum fluctuations. In topological confinement, instead, electric charge is linearly confined and the photon becomes a massive antisymmetric tensor via the St\"uckelberg mechanism. Oblique confinement phases arise when the string condensate carries both
magnetic and electric flux (dyonic strings). Such phases are characterized by a vortex quantum Hall effect potentially relevant for the dissipationless transport of information stored on vortices. 
\end{abstract}

\maketitle

\section{Introduction}
In the last twenty years it has been realized that quantum degrees of freedom can organize themselves into very robust, gapped, emergent macroscopic states independently of any spontaneous symmetry breaking pattern. This new type of order has been called topological order  \cite{Wen} since it is characterized by typical degeneracies on topologically non-trivial manifolds. 

One of the hallmarks of topological order is that the low-energy effective field theory for such states involves a topological field theory \cite{birmi}, i.e. a background-independent quantum field theory. 
The general idea \cite{Wen} is to write the conserved matter current describing the fluctuations about topologically ordered states as the curl of a gauge field and to formulate the effective field theory for the fluctuations as a gauge theory. In two spatial dimensions, the dominant gauge action at long distances is the famed Chern-Simons term \cite{jackiw, witten}, which describes incompressible quantum Hall fluids and their edge excitations.

Recently, a new topological state of matter in three dimensions has attracted much interest. This is the so-called topological insulator \cite{topins}, realized in materials which are insulating in the bulk but have gapless metallic edge states. As we have shown in \cite{dst1} \cite{dst2} this state is characterized by a low-energy effective field theory containing the topological BF term \cite{birmi}, which is  a $P$- and $T$-conserving, higher-dimensional generalization of the Chern-Simons term, and can be realized "synthetically" in fabricated arrays of Josephson junctions. The relevance of the BF term as an effective low-energy description of topological insulators has been recently re-emphasized in \cite{moo}. 

The BF term, however, is not the only possible marginal term in the effective low-energy action in 3D: a conventional Maxwell term must be added to fully determine the possible phases of topological matter. The phase structure is governed by three parameters: the topological BF coupling and the electric permittivity and the magnetic permeability of the topological insulator \cite{dst2}. These parameters drive the condensation of topological defects associated with the compactness of the $U(1)$ gauge symmetries  characterizing the topological order. The condensation of electrically charged solitons for high electric permittivity and low magnetic permeability is the origin of topological superconductivity, the condensation of magnetic vortex strings for low electric permittivity and high magnetic permeability causes the material to undergo a transition to a new state of matter characterized by charge confinement. 

There are two types of topological insulators: normal (or weak) topological insulators and strong topological insulators. In the latter the magnetic vortex strings carry also electric flux (dyons) \cite{dst2}, giving rise to a quantized $\theta$-term \cite{jackiw2} in the effective action for the edge states in presence of time-reversal-breaking perturbations. This $\theta$-term is the origin of the anomalous magneto-electric response and the magneto-optical properties of strong topological insulators  \cite{topins}. 

In this paper we shall derive the effective electromagnetic action induced by the condensation of the topological defects in the these new phases of matter. It is well known that, in conventional superconductors, photons acquire a mass through spontaneous symmetry breaking and the Anderson-Higgs mechanism. What is the corresponding effective action in topological superconductors? And what is the fate of the photon in the topological confinement phase? We shall show that in topological superconductors the photon acquires itself a topological mass via the BF mechanism \cite{bow}. There is no Higgs fields, the degree of freedom that is "eaten" by the photon is a scalar mode due to the condensation of electric solitons. The topological confinement phase is dual to topological superconductivity: in this case it is the new degree of freedom arising from the condensation of magnetic vortex strings that "eats" the original photon via the StŸckelberg mechanism \cite{stu}. This mechanism turns the photon into a massive antisymmetric tensor that couples to electric strings between charge-anticharge pairs, thus realizing linear charge confinement. The dual Meissner effect confines electric fields into thin tubes between charges: the motion of these electric strings is governed by a tensor version of the London equations. In oblique confinement phases, with a dyonic vortex condensate, the $\theta$-term of strong topological insulators induces a BF term in the bulk of the material. As we will show, this induces a new physical effect which is one of the main results of this paper. This effect is a 3D analogue of the 2D quantum Hall effect and represents a quantum Hall effect for vortices.

In order to derive these results we shall make use of an old idea of Julia and Toulouse \cite{jt}. The insight of Julia and Toulouse is that the condensation of  topological defects in solid state media 
generates new hydrodynamical modes for the low-energy effective theory: these new modes 
are essentially the long wavelength fluctuations of the continuous distribution of topological 
defects. Moreover, Julia and Toulouse proposed also a generic prescription to identify 
these new modes. However, in the framework of ordered solid-state media, it is difficult 
to write down an action for the phase with a condensate of topological defects due to the 
non-linearity of the topological currents, the lack of gauge invariance, and the need to 
introduce dissipation terms. On the other side, the Julia-Toulouse prescription is sufficient to fully determine the low-energy action due to the condensation of topological defects in generic compact antisymmetric field theories, for which none of the above problems is 
present \cite{que}. This is exactly what is needed to determine the effective electromagnetic actions of topological matter. 

For ease of presentation of the key principles we shall use relativistic notation in Euclidean space-time, and units in which $\hbar =1$, $c=1$. 

This paper is organized as follows. In Section II we give  a brief introduction to the relevant topological field theory describing 3D topological states of matter. In Section III we present the results for the induced action in  the electric and magnetic condensation phases, in particular we show how $U(1)$ confinement phase arises via the St\"uckelberg  mechanism. We show also how a vortex Quantum Hall effect takes place if the condensed magnetic vortices also carry electric field. Section IV is devoted to a few concluding remarks.

\section{3D topological matter}

The topological BF action is the field theory description of 3D Abelian topological states of matter.
It can be defined in any space-time dimensions $(D+1)$ \cite{birmi} as the wedge product of a p-form $a$  and the curvature $db$ of a (D-p) form $b$:
\begin{equation}
S_{\rm BF} = {k \over 2 \pi} \int_{M_{D+1}}  a_p \wedge d b_{D-p}\ ,
\label{abf}
\end{equation}
where $k$ is a dimensionless coupling constant.
This action has a generalized Abelian gauge symmetry under the transformation 
\begin{equation}
b \rightarrow b + \eta \quad , a \rightarrow a + \xi \ ,
\nonumber
\end{equation}
where $\eta$ and $\xi$ are a closed p and a closed (D-p) form, respectively. If we add a kinetic term for the $a_p$ form and the $ b_{D-p}$ form,  we see that the BF term is the generalization to any number of dimensions of the  Chern-Simons mechanism for topological mass generation:
\begin{eqnarray}S_{TM} =  \int_{M_{D+1}} &&{k \over 2 \pi} a_p \wedge d b_{D-P}  - {1 \over 2 e^2} d a_p \wedge * d a_p + \nonumber \\ 
&&+ {(- 1)^{D-p} \over 2 g^2} d b_{D-p}\wedge * d b_{D-p} \ ;
\label{topmas}
\end{eqnarray}
which describes a theory with topological mass given by $m = { k e g \over 2 \pi}$.This topological mass  plays the role of the gap characterizing topological states of matter.

For applications to condensed matter systems we are considering $a$  a 1-form, namely $p =1$. Moreover, since $[g^2] = m^{D-1}$, in $D = 3$ the kinetic term for the $b_2$ form is irrelevant and
the low-energy effective action for topological matter \cite{dst1} \cite{dst2} is written in terms of two gauge fields, a one-form $a_{\mu}$ and a two-form $b_{\mu \nu}$ as: 
\begin{equation}
S = {ik \over 4 \pi} \int d^4x \ b_{\mu \nu} \epsilon_{\mu \nu \alpha \beta} f_{\alpha \beta} +  
{1\over 4 e^2 } \int d^4x \ f_{\mu \nu} f_{\mu \nu} \ .
\label{one}
\end{equation}
Here $f_{\mu \nu} = \partial_{\mu}a_{\nu}-\partial_{\nu}a_{\mu}$ is the field strength associated with $a_{\mu}$ and $k$ (the BF coupling) and $e$ are dimensionless couplings. 
The idea is that 
\begin{eqnarray}
j_{\mu} &&= {k\over 2\pi} \epsilon_{\mu \nu \alpha \beta}\partial_{\nu}b_{\alpha \beta} \ ,\nonumber \\
\phi_{\mu \nu} &&= {k\over 32\pi^2} \epsilon_{\mu \nu \alpha \beta}\partial_{\alpha} a_{\beta} \ ,
\label{two}
\end{eqnarray}
are the conserved charge and magnetic flux currents representing the low-energy fluctuations about a topologically ordered state and (\ref{one}) is the most general gauge invariant marginal action for these degrees of freedom. 

The action (\ref{one}) has two $U(1)$ gauge symmetries under the transformations:
\begin{eqnarray}
a_{\mu} &&\to a_{\mu} + \partial_{\mu} \xi \ , \nonumber \\
b_{\mu \nu} &&\to b_{\mu \nu} + \partial_{\mu}\xi_{\nu} - \partial_{\nu}\xi_{\mu} \ .
\label{three}
\end{eqnarray}
The important point is that, for topologically ordered states,  these gauge symmetries are compact and the compactness of the $U(1)$ gauge group, as usual, leads to the presence of topological defects. As is beautifully explained in \cite{polbo}, the compactness of  $U(1)$ gauge groups inevitably introduces a mass scale, the compactification radius of the gauge fields. Typically, this UV cutoff arises from spontaneous symmetry breaking from  a larger, non-Abelian compact group down to $U(1)$ on a given mass scale $m$. From the point of view of the $U(1)$ effective field theory on energy scales much smaller than $m$, the topological defects can be viewed essentially as singularities of dimension $(d-h-1)$ in $R^{d+1}$, where $h$ is the degree of the associated antisymmetric form \cite{que}. In the present case this amounts to point-like electric solitons associated with the two-form $b_{\mu \nu}$ and string-like magnetic vortices associated with the one-form $a_{\mu}$. The low-energy effective action is then well-defined only outside these singularities, which contain lumps of energy (action) involving higher-lying fields. 

For topological matter the relevant singularities arise from the compactness of the BF term \cite{dst1} \cite{dst2} and are due to the periodicities of the gauge fields in this term, exactly as in the well-known example of compact QED \cite{polbo}. They can be taken into account in the action by coupling the gauge fields to currents representing the world-lines of point-like electric solitons and the world-surfaces of infinitely thin magnetic vortices. 
\begin{equation}
S \to S +  i k \int d^4 x \ a_{\mu} J_{\mu} + b_{\mu \nu} \Phi_{\mu \nu} \ ,
\label{four}
\end{equation}
where the singular currents are such that
\begin{equation}
\int_{S^3} d^3 x \ J_0 = {\rm integer} \ , 
\int_{S^2} d^2 x \ n_i \Phi_{0i} = {\rm integer} \ ,
\label{five}
\end{equation}
where $S^3$ is any three-sphere containing the electric singularity and $S^2$ is any two-sphere on a plane perpendicular to the unit direction ${\bf n}$ of the flux vortex. 

In order to derive the effective action for the electromagnetic field in topological matter we couple the charge and flux modes to it,
\begin{equation}
S \to S + \int d^4 x \ i e A_{\mu}j_{\mu} + i \phi  F_{\mu \nu}\phi_{\mu \nu} \ ,
\label{six}
\end{equation}
where $e$ is the electron charge and the second coupling represents a time-reversal-breaking term due to possible electric field-carrying dyonic vortices. A Gaussian integration over the matter modes $a_{\mu}$ and $b_{\mu \nu}$ gives the electromagnetic effective action in the presence of topological defects. This Gaussian integration is best performed by adding an infrared irrelevant kinetic term
\begin{equation}
S_{\rm reg} = {1\over 12 g^2} \int d^4 x \ h_{\mu \nu \alpha}h_{\mu \nu \alpha} \ ,
\label{seven}
\end{equation}
as an ultraviolet regulator. Here $h_{\mu \nu \alpha} =  \partial_\mu b_{\nu \alpha} + \partial_\nu b_{ \alpha \mu} +\partial_\alpha b_{\mu \nu }$ is the field strength associated with the two-form gauge field $b_{\mu \nu}$ and $g$ is a coupling with dimension mass. This regulator term makes the quadratic kernels well defined by inducing a mass $m=egk/\pi$ for all fields. This mass can then be removed again after the integration by letting $g \to \infty$, giving the result
\begin{eqnarray}
&&S_{\rm eff} = \int d^4x \ {1\over 4} \left( F_{\mu \nu}- {4\pi \over e}\Psi_{\mu \nu} \right) \left( F_{\mu \nu}-{4\pi \over e} \Psi_{\mu \nu} \right) \nonumber \\
&&+ \int d^4x \ \left( i{\theta \over 32\pi^2} F_{\mu \nu} - i e k I_{\mu \nu} \right) \epsilon_{\mu \nu \alpha \beta}
\left( F_{\alpha \beta} - {4\pi \over e} \Psi_{\alpha \beta} \right) \ ,
\label{eight}
\end{eqnarray}
where $J_{\mu} = 2 \epsilon_{\mu \nu \alpha \beta} \partial_{\nu} I_{\alpha \beta}$, $\Psi _{\mu \nu}= \epsilon_{\mu \nu \alpha \beta} \Phi_{\alpha \beta}$ and $\theta = ke\phi /2$. When the topological excitations satisfy the Dirac quantization condition $ke\phi /2\pi = {\rm integer}$ the $\theta$ parameter becomes quantized in integer units of $\pi$, $\theta = n\pi$, $n\in  \mathbb{ Z}$ and time-reversal symmetry is restored. 

\section{Effective action induced by the condensation of topological defects}

At this point we are ready to apply the Julia-Toulouse prescription \cite{jt} \cite{que} to derive the effective action in the topological phases. Let us begin with the topological superconductor phase, in which the magnetic vortices are dilute, whereas the electric solitons condense. In this case we can set $\Psi_{\mu \nu} = 0$, whereas the proliferating singularities carrying the delta-like current densities $I_{\mu \nu}$ get promoted to a continuous two-form antisymmetric field $B_{\mu \nu}$ defined on the whole space, without singularities: $ -i I_{\mu \nu} \to B_{\mu \nu}$. 
The kinetic term for these new degree of freedom is given to lowest order by the square of the three-form field strength $H_{\mu \nu \alpha} =  \partial_\mu B_{\nu \alpha} + \partial_\nu B_{ \alpha \mu} +\partial_\alpha B_{\mu \nu }$: 
\begin{eqnarray}
S_{\rm eff}^{TS} &&= \int d^4 x \ {i \pi k} \ B_{\mu \nu}  \epsilon_{\mu \nu \alpha \beta}  F_{\alpha \beta} + {1\over 4} F_{\mu \nu} F_{\mu \nu} \nonumber \\
&&+ \int d^4x \ {1\over 12\Lambda ^2} H_{\mu \nu \alpha} H_{\mu \nu \alpha} \ ,
\label{nine}
\end{eqnarray}
where we have used gauge invariance to reabsorb a possible $\theta$-term into the new two-tensor degree of freedom and $\Lambda$ is a new mass scale describing, essentially, the average density of the condensed charges. 
This action describes a topologically massive photon \cite{bow} with quantized mass $m=4 \pi k\Lambda $. Gauge invariance insures that the antisymmetric Kalb-Ramond field \cite{birmi} embodies a single scalar degree of freedom: it is this scalar degree of freedom that is "eaten" by the original photon which thereby becomes massive. The Kalb-Ramond scalar, however is in no way tied to any symmetry breaking mechanism; it arises only as a consequence of quantum mechanical condensation of topological excitations, in analogy to the famed Kosterlitz-Thouless transition \cite{kt} or the confining transition in 3D QED \cite{polbo}. This is the mechanism of topological superconductivity \cite{bow}.

Let us now proceed to the dual situation, in which electric solitons are dilute whereas magnetic vortices condense. In this case it is the magnetic singularities embedded in $\Psi_{\mu \nu}$ that get promoted,  to a new continuous antisymmetric form $B_{\mu \nu}$ on the whole space, without singularities, $(4\pi/e) \Psi_{\mu \nu} \to B_{\mu \nu}$, giving
\begin{eqnarray}
S_{\rm eff}^{TC} &&= \int d^4x \  {i\theta \over 32 \pi^2} F_{\mu \nu}  \epsilon_{\mu \nu \alpha \beta} \left( F_{\alpha \beta} - B_{\alpha \beta} \right) \nonumber \\
&&+\int d^4x  {1\over 4} \left( F_{\mu \nu}-B_{\mu \nu} \right) \left( F_{\mu \nu}-B_{\mu \nu} \right) \nonumber \\
&&+ {1\over 12\Lambda ^2} H_{\mu \nu \alpha} H_{\mu \nu \alpha} \ .
\label{ten}
\end{eqnarray}
The original field strength $F_{\mu \nu}$ can now be reabsorbed into the new field $B_{\mu \nu}$ by a gauge transformation ($\ref{three}$): $B_{\mu \nu} \to B_{\mu \nu} + F_{\mu \nu}$. 
The electromagnetic effective action becomes thus
\begin{eqnarray}
S_{\rm eff}^{TC} &&= \int d^4x \  {i\theta \over 32\pi ^2} B_{\mu \nu}  \epsilon_{\mu \nu \alpha \beta} F_{\alpha \beta}
+{1\over 4}  B_{\mu \nu} B_{\mu \nu} \nonumber \\
+&& {1\over 12\Lambda ^2} H_{\mu \nu \alpha} H_{\mu \nu \alpha} \ .
\label{thirteen}
\end{eqnarray}
In exact opposition to what is happening in topological superconductors, here it is the new degrees of freedom arising from the condensation of topological defects that "eat" the original photon to become a massive ($m=\Lambda$) two-form Kalb-Ramond field, which also contains three independent massive degrees of freedom. This is the St\"uckelberg mechanism \cite{stu} which is dual to the Higgs mechanism. Also in this case no symmetry breaking is involved, the new degrees of freedom arise purely from the quantum condensation of topological defects. 

The question arises as to what exactly characterizes this new phase of matter. This question can be readily answered by calculating the Wilson loop order parameter with the effective action ($\ref{thirteen}$). The Wilson loop is the typical (non-local) order parameter which characterizes the behavior of gauge theories. It is defined as 
\begin{equation}
W(C) = < {\rm exp} \ i\int_C dx^{\mu}A_{\mu} > \ , 
\label{fourteen}
\end{equation}
where $C$ denotes a closed loop. Actually, the St\"uckelberg mechanism implies that the massive Kalb-Ramond field couples to antisymmetric tensor currents $\omega_{\mu \nu}$ such that $2 \partial _{\nu} \omega_{\mu \nu} =  j_{\mu}$, where $ j_{\mu}$ is the original charged matter current. This implies that the Wilson loop is turned by the St\"uckleberg mechanism into a Wilson surface order parameter
\begin{equation}
W(S) = < {\rm exp} \ i\int_S dx^{\mu} \wedge dx^{\nu} B_{\mu \nu} > \ , 
\label{fifteen}
\end{equation}
where $S$ is a surface whose boundary is $C$. The mass term for the Kalb Ramond field then implies the area-law
\begin{equation}
W(S) = {\rm exp}\left( -TA(S) \right) \ ,
\label{sixteen}
\end{equation}
where $A(S)$ is the area of the surface $S$ and the string tension $T$ is determined by the new mass scale as $T=\Lambda^2 K_0(\Lambda \epsilon )/4\pi $ where $K_0$ is a Bessel function and $\epsilon$ is the ultraviolet cutoff \cite{que}. An area-law for the Wilson loop order parameter implies a linear potential between charges, which is tantamount to confinement: electric fields cannot penetrate the medium and are squeezed into thin tubes (strings) that cause the charges at the end of these tubes to be  linearly bound into neutral pairs. The tensor current of such electric flux tubes is exactly the field $\omega_{\mu \nu}$ coupling to the Kalb-Ramond gauge field. 
This is one of the most important results in this paper: topological matter with a compact BF term can realize $U(1)$ confinement via the St\"uckelberg mechanism. 

Just as the Meissner effect implies the London equation ${\bf j}^{\rm ind} \propto {\bf A}$
for the induced current in superconductors, confinement implies a corresponding "tensor London equation" for the electric flux tubes in the confinement phase. The induced electric flux tube current in this phase can be obtained as the functional derivative of the effective action with respect to $B_{\mu \nu}$. The dominant part of this current  (for $\theta = 0$) is given by 
\begin{equation}
\omega^{\rm ind} _{\mu \nu} = \  {\delta \over \delta B_{\mu \nu}} S_{\rm eff}^{TC} = {1\over 4} \ B_{\mu \nu}
\label{seventeen}
\end{equation}
which translates into the following gauge invariant equations 
\begin{equation}
\tilde H_{\mu} = 2 \epsilon_{\mu \nu \alpha \beta} \partial_{\nu} \omega^{\rm ind}_{\alpha \beta} \  .
\label{eighteen}
\end{equation}

A very interesting new effect arises in confinement phases, in which the condensed magnetic vortices also carry electric field. Such confinement phases with a dyonic vortex condensate are customarily called "oblique confinement phases" and they arise as "daughters" of strong topological insulators. 
In such phases there is an additional BF term in the effective electromagnetic action, which is the descendant of the original $\theta$-term of the parent strong topological insulator and arises from a direct coupling of the vortex current $\phi_{\mu \nu}$ to the new gauge tensor field $B_{\mu \nu}$ \cite{dst2}. This BF term represents a vortex quantum Hall effect for strings that carry both magnetic and electric flux. It can be equivalently characterized by one of the following two equations for the induced magnetic and electric flux currents
\begin{eqnarray}
\phi^{\rm Hall} _{ij} &&=  {\theta \over 16 \pi^2} \epsilon_{ijk} E_k \ ,
\nonumber \\
\omega^{\rm Hall} _{ij} &&=  {\theta \over 16 \pi^2} \epsilon_{ijk} B_k \ .
\label{nineteen}
\end{eqnarray}
An external electric field induces a magnetic vortex current perpendicular to both the applied electric field and the direction of the magnetic flux. An external magnetic field induces an electric flux tube current perpendicular to both the applied magnetic field and the direction of the electric flux. 
The vortex Hall conductivity is quantized: $\sigma_H = n/ 16\pi$, $n\in \mathbb{Z}$. It is conceivable that this vortex quantum Hall effect might find an application for the dissipationless transport of information stored on dyonic vortices in oblique confinement phases.  

\section{Conclusions}
In this paper we have used the Julia-Toulouse mechanism to derive the effective action for the electromagnetic field in the topological superconductor and topological confinement phase of 3D topological states of matter characterized by the presence in their action of a topological BF term. This allowed us to derive what is the fate of the photon in these new states of matter.  In  conventional superconductors, photons acquire a mass through spontaneous symmetry breaking. In the topological superconducting phase, instead, mass  arises as a consequence of quantum mechanical condensation of topological excitations.
The antisymmetric Kalb-Ramond field, that embodies a single scalar degree of freedom, is "eaten" by the original photon that become thus massive. In the confinement phase instead, is the new degrees of freedom arising from the condensation of topological defects that "eats" the original photon to become a massive two-form Kalb-Ramond field: this is the St\"uckelberg mechanism which is dual to the Higgs mechanism. No spontaneous symmetry breaking is required in these topological mechanisms.

In the  oblique confinement phases,  in which the condensed magnetic vortices also carry electric field an interesting effect occurs: an 
 additional BF term is present in the effective electromagnetic action.  This is the descendant of the original $\theta$-term of the parent strong topological insulator, and
gives rise to a vortex quantum Hall effect for dyonic strings.

\end{document}